\def\be{\begin{equation}}
\def\ee{\end{equation}}
\def\ba{\begin{array}}
\def\ea{\end{array}}

\documentclass[aps,amsmath,amssymb,amsfonts,pra]{revtex4}
\usepackage{graphicx}
\usepackage{epstopdf}
\def\qed{\leavevmode\unskip\penalty9999 \hbox{}\nobreak\hfill
     \quad\hbox{\leavevmode  \hbox to.77778em{%
               \hfil\vrule   \vbox to.675em%
               {\hrule width.6em\vfil\hrule}\vrule\hfil}}
     \par\vskip3pt}

\newtheorem{definition}{Definition}
\newtheorem{theorem}{Theorem}
\newtheorem{corollary}{Corollary}
\newtheorem{lemma}{Lemma}
\begin{document}

\title{Classifying coherence with a finite set of witnesses}

\author{Xue-Na Zhu$^{1}$}
\thanks{jing\_feng1986@126.com}
\author{Zhi-Xiang Jin$^{2}$}
\thanks{jzxjinzhixiang@126.com}
\author{Gui Bao$^{1}$}
\thanks{baoguigui@163.com}
\author{Shao-Ming Fei$^{3}$}
\thanks{feishm@cnu.edu.cn}
\affiliation{$^1$School of Mathematics and Statistics Science, Ludong University, Yantai 264025, China\\
$^2$School of Computer Science and Technology, Dongguan University of Technology, Dongguan, 523808, China\\
$^3$School of Mathematical Sciences, Capital Normal University, Beijing 100048, China}

\begin{abstract}
Coherence is a fundamental resource in quantum information processing, which can be certified by a coherence witness.
In order to detect all the coherent states, we introduce  a useful concept of coherence witness and structure the set of  coherence witnesses $C^{d}_{[m,M]}$.
We present necessary and sufficient conditions of detecting quantum coherence of $d-$dimensional quantum states based on $C^{d}_{[m,M]}$.
Moreover,
we show that each coherent state can be detected by one of the coherence witnesses in a finite set. The corresponding finite set of coherence witnesses is presented explicitly, which detects all the coherent states.
\end{abstract}

\maketitle

\section{Introduction}
Coherence is a vital physical resource with various applications in such as
biology \cite{1,2,3}, thermodynamical systems \cite{4,5}, transport theory \cite{6,7} and nanoscale physics \cite{8}. Coherence is also tightly related to quantum correlations and can be transformed into quantum entanglement and nonlocality under incoherent operations \cite{zhuhj,xiya}.
It can be used to test the
existence of quantumness in a single system and quantum
correlations in composite systems and has many applications\cite{KDW,YD}.
To quantify the coherence of quantum states from the perspective of the resource theory, a rigorous framework has been proposed in \cite{CX,CS,FL,IM,DG,J}.
Within such a framework various coherence measures have been presented, such as the relative
entropy coherence, the $l_1$-norm coherence \cite{CX}, the experimentally friendly measure of coherence based on the Wigner-Yanase-Dyson skew information\cite{DG},
the measures based
on entanglement\cite{AS},
the measure based on robustness\cite{prl116,93}, and so on\cite{YT,WZ,YQ,SA,TR}.
Some coherence
measures have analytical results. But there also exist coherence
measures which do not have analytical expressions yet.
Thus, it is an interesting problem to estimate coherence measures,
especially for experimental states\cite{ZM}.

An equivalent approach of identifying coherence is
based on special types of Hermitian operators, called coherence witnesses.
In contrast to conventional
methods that rely on state tomography, coherence
witnesses can directly identify the coherent states\cite{ZM}.
Coherence witnesses, like the entanglement witnesses
\cite{TTY,JW,JW2,BH,YC,JH,LF,CG}, have emerged as potent tools for coherence
detection in experimental settings and coherence quantification in theoretical contexts.
One of experimentally implementable ways to detect coherence is to measure the expectation value of a coherence witness.
Several experiments relevant to coherence witness have been reported\cite{93,YQ,WZ,MRTR}.
The
coherence witness was first introduced in Ref.\cite{prl116}.
A Hermitian operator $W$ is a coherence witness if and only if $Tr(W\delta)\geq0$ for all incoherent states $\delta$, and there exists at least a coherent state $\rho$ such that $Tr(W\rho)<0$,
which imply that all the
diagonals of $W$ are non-negative, and  $W$ has some negative eigenvalues\cite{prl116}.
Refs.\cite{ZHD,BHW,HR} proposed the stringent coherence witness such that (C1) $Tr(W\delta)=0$; and (C2) $Tr(W\rho)\not=0$.
The strengthened
conditions (C1) and (C2) imply that all the diagonals of $W$ must be
zero.
Ref.\cite{ZM} proposed the coherence witness
$W=\sum_{i=1}^{d}|i\rangle\langle i|\rho|i\rangle\langle i|-\rho$,
where its diagonal elements are zero and
off-diagonal elements are complex numbers with modulus 1.
More recently, the set of all coherence
witnesses with the same trace is defined by Ref.\cite{pra109},
$W_R=\{W\in \emph{H}_{\geq}|Tr(W)=R\}$, where $\emph{H}_{\geq}$ is  the
set of $d\times d$ Hermitian matrices with non-negative diagonal
elements.

We notice that those coherence
witnesses in Refs.\cite{prl116,ZHD,BHW,HR,ZM,pra109} are defined based on  non-negative diagonal
elements.
In this work,
we introduce a useful concept of coherence witness $W_{m}^{M}$
based on the fixed minimum value $m$ and maximum value $M$ of the diagonals of Hermitian operator.
The main differences
between the coherence witnesses in Refs.\cite{prl116,ZHD,BHW,HR,ZM,pra109} and
$W_{m}^{M}$ are that the diagonals of $W_{m}^{M}$ may be positive or negative.
Furthermore, we demonstrate that knowing the maximum and minimum
values  of the diagonals for a given observable indeed enhances
our ability to detect coherent. We establish a finite set of coherence witnesses, which detects all the coherent states.

\section{Witnessing quantum coherence with a finite set of witnesses}
Let $H$ be a finite $d$-dimensional Hilbert space with fixed   orthonormal
basis $\{|i\rangle\}^{d-1}_{i=0}$. An incoherent state $\delta$ in this basis is of the following form,
\begin{equation}
\delta=\sum_{i=0}^{d-1}p_i|i\rangle\langle i|,
\end{equation}
where $p_i$ are probabilities. We denote $\mathcal{I}$ the set of all incoherent states. Denote $\mathcal{D}$ the set of all density matrices. All the states within $\mathcal{D} \setminus \mathcal{I}$ are referred to coherent ones \cite{CX}.

In the following for each real number $m\leq M$, we denote $W^{M}_{m}=[(W^{M}_{m})_{ij}]_{d\times d}$ a $d\otimes d$ Hermitian matrix such that $\min_{i}\{(W^{M}_{m})_{ii}\}=m$ and $\max_{i}\{(W^{M}_{m})_{ii}\}=M$.

\begin{lemma}\label{Le1}
If $\delta$ is an incoherent state, then $Tr(W^{M}_{m}\delta)\in[m,M].$
\end{lemma}

{\sf [Proof]}~For any incoherent states $\delta=\sum_{i=0}^{d-1}p_i|i\rangle\langle i|,$
one has $Tr(W^{M}_{m}\delta)=\sum_{i=1}^{d}(W^{M}_{m})_{ii}p_{i-1}$. In fact,
$m=\sum_{i=1}^{d} m p_{i-1}\leq\sum_{i=1}^{d}(W^{M}_{m})_{ii}p_{i-1}\leq \sum_{i=1}^{d} M p_{i-1}=M$ due to that
$m=\min_i\{(W^{M}_{m})_{ii}\}\leq (W^{M}_{m})_{ii}\leq\max_i\{(W^{M}_{m})_{ii}\}=M$, $p_i\in[0,1]$ and $\sum_{i=1}^{d}p_{i-1}=1$. Therefore, we have $m\leq Tr(W^{M}_{m}\delta)\leq M$. $\Box$

Lemma \ref{Le1} motivates us to define the set of all coherence witnesses with the same maximum value and minimum value of the diagonals. Denote $\mathcal{R}=(-\infty,+\infty)$ and $\Delta\subset \mathcal{R}$ a given non-empty real set, such that $\mathcal{R}\backslash\Delta\neq{\O}$.

\begin{definition}\label{D1}
For any given $\Delta$, we say that a $d\times d$ Hermitian operator $W$ is a coherence
witness if (1) for each incoherent state $\delta$, $Tr(W\delta)\in \Delta$; (2) there exits at least a coherent state $\rho$ such that $Tr(W\rho)\in\mathcal{R}\backslash\Delta$.
\end{definition}

Definition \ref{D1} actually defines a family of coherence witnesses which includes several typical ones. i) $\Delta=[0,+\infty)$, $W$ is the coherence witness in Ref. \cite{prl116}. ii)  $\Delta=0$, $W$ is the coherence witness of Refs. \cite{HR,BH,ZM}. iii) $\Delta=[0,Tr[W]]$,  $W$ is the coherence witness in Ref. \cite{pra109}.

According to the Lemma \ref{Le1} we set $\Delta=[m,M]$.
Denote $C^{d}_{[m,M]}=\{W^{M}_{m}=[(W^{M}_{m})_{ij}]_{d\times d}|W^{M}_{m} ~are~coherence~witnesses~of~\rho\in \mathcal{D} \setminus \mathcal{I}\}$.

\begin{lemma}\label{Le2}
$C^{d}_{[m,M]}$ is a nonempty set of Hermitian operators.
\end{lemma}

{\sf [Proof]}~
Consider the following Hermitian operator,
\begin{eqnarray*}
W&=\begin{pmatrix}
M&\frac{d-m+M}{2}&0&0&...&0&0\\
\frac{d-m+M}{2}&m&0&0&...&0&0\\
0&0&M&0&...&0&0\\
0&0&0&M&...&0&0\\
.&.&.&.&...&.&.\\
.&.&.&.&...&.&.\\
0&0&0&0&...&M&0\\
0&0&0&0&...&0&M\\
\end{pmatrix}.
\end{eqnarray*}
It is noticeable that $\min_i\{W_{ii}\}=m$ and $\max_i\{W_{ii}\}=M$.
Consider the coherent state, $\rho=\frac{1}{d}(|0\rangle\langle 1|+|1\rangle\langle 0|)+\sum_{i=0}^{d-1}\frac{1}{d}|i\rangle\langle i|.$
Obviously, $Tr(W\rho)=M+1>M\geq m$. i.e., $Tr(W\rho)\in\mathcal{R}\backslash[m,M]$.
According to Lemma 1 and Definition \ref{D1}, $W$ is a coherence witness of $\rho$,
i.e., $W\in C^{d}_{[m,M]}$. Hence, $C^{d}_{[m,M]}$ is not an empty set.$\Box$

Now consider a coherence witness $W^M_m\in C^{d}_{[m,M]}$ such that $(W^M_m)_{11}=m$ and $(W^M_m)_{22}=M$ without loss of generality. From Lemma \ref{Le1},
we have $m\leq Tr(W^M_m\delta)\leq M$ for each incoherent state $\delta$. On the other hand, there exists an incoherent state $\delta_{h}$ such that $Tr(W^M_m\delta_h)=h$ for any
$h\in[m, M]$, where $\delta_{h}=\sum^{d-1}_{i=0}p_i|i\rangle\langle i|$ with
\begin{equation*}
\left\{
\begin{aligned}
&p_i=\frac{1}{d},~for~i=0,1,...,d-1,~~~~~~~~~~~~~~~~~~~~~~~~~~~~~~~~~~~~~~if~m=M;\\
&p_{0}=\frac{M-h}{M-m},~p_{1}=\frac{h-m}{M-m},~p_i=0~for~i=2,3,...,d-1,~~if~m< M.
\end{aligned}
\right.
\end{equation*}

\begin{theorem}\label{TH1} A state $\rho$ is coherent if and only if there is a $W\in C^{d}_{[m,M]}$ such that either $Tr(W\rho)<m$ or $Tr(W\rho)>M$.
\end{theorem}

{\sf [Proof]}~$(\Rightarrow)$~~Suppose $\rho$ is a coherence state. Without loss of generality, we assume $\rho_{kl}=\langle k|\rho|l\rangle\not=0$ for $k\not=l$. We define $W^{Re}_{k,l}=\frac{1}{2}(|k\rangle\langle l|+|l\rangle\langle k|)$ and
$W^{Im}_{k,l}=\frac{i}{2}(|k\rangle\langle l|-|l\rangle\langle k|).$

{\sf Case 1} $m=M$. We need to prove that there is a $W\in C^{d}_{[m,m]}$ such that $Tr(W\rho)\not=m.$ Let
\begin{equation*}
W=\left\{
\begin{aligned}
&W^{Re}_{k,l}+mI_{d},~~~~Re(\rho_{kl})\not=0;\\
&W^{Im}_{k,l}+mI_{d},~~~~Re(\rho_{kl})=0,
\end{aligned}
\right.
\end{equation*}
where $Re(z)=\alpha$ and $ Im(z)=\beta $
for any complex number $z=\alpha + \beta i \in \mathcal{C}$, and $I_{d}$ denotes the $d\times d$ identity matrix.
We have
\begin{equation*}
Tr(W\rho)=\left\{
\begin{aligned}
&Re(\rho_{kl})+m,~~~~Re(\rho_{kl})\not=0;\\
&Im(\rho_{kl})+m,~~~~Re(\rho_{kl})=0,
\end{aligned}
\right.
\end{equation*}
i.e., $Tr(W\rho)\not=m.$

{\sf Case 2} $m<M$. According to Lemma \ref{Le2}, we take a coherence witness $W^{M}_{m}\in C^{d}_{[m,M]}$. Let
\begin{equation*}
W=\left\{
\begin{aligned}
&\varepsilon_{Re}W^{Re}_{k,l}+W^{M}_{m},~~~~Re(\rho_{kl})\not=0;\\
&\varepsilon_{Im}W^{Im}_{k,l}+W^{M}_{m},~~~~Re(\rho_{kl})=0,
\end{aligned}
\right.
\end{equation*}
where $\varepsilon_{Re}=\frac{M+1-Tr(W^{M}_{m}\rho)}{Re(\rho_{kl})}$ with $Re(\rho_{kl})\not=0$ and $\varepsilon_{\emph{Im}}=\frac{M+1-Tr(W^{M}_{m}\rho)}{Im(\rho_{kl})}$ with $Im(\rho_{kl})\not=0$.
As $\rho_{kl}\not=0$, we have
$Tr(W\rho)=M+1> M$.

From cases 1 and 2 we have that there exists a $W\in C^{d}_{[m,M]}$
which detects a given coherent state $\rho\in\mathcal{D} \setminus \mathcal{I}$.

$(\Leftarrow)$~~If there is a $W\in C^{d}_{[m,M]}$ such that $Tr(W\rho)<m$
or $Tr(W\rho)>M$, $\rho$ is coherent according to Lemma \ref{Le1}. $\Box$

We denote $W(\rho)=\{\rho|Tr(W\rho)\in \mathcal{R}\backslash\Delta\}$
and $B(\rho)=\{W_1(\rho),W_2(\rho),...,W_{k}(\rho)\}$, where $B=\{W_i\}_{i=1}^k$ is a set of $d\times d$ Hermitian operators. We say that two sets of Hermitian operators $B_1$ and $B_2$ are equivalent if $B_1(\rho)=B_2(\rho)$. From the proof of Theorem \ref{TH1}, we immediately verify that each coherent state can be witnessed by at least one $W^M_m\in C^{d}_{[m,M]}$ and $C^{d}_{[m,M]}(\rho)=\mathcal{D} \setminus \mathcal{I}$.
Furthermore, we want to find a finite nonempty set
$\widetilde{C^{d}}_{[m, M]}\subset C^{d}_{[m,M]}$, which satisfies
$\widetilde{C^{d}}_{[m, M]}(\rho)=C^{d}_{[m,M]}(\rho)$.

\noindent{\it Example 1} Consider a qubit state, $\rho=\frac{1}{2}(I_2+x\sigma_1+y\sigma_2+z\sigma_3)$, where $x^2+y^2+z^2\leq1$ and $\sigma_i$ (i=1,2,3) are the standard Pauli matrices. Let $W^{[K,a,b,c]}=\frac{1}{2}(KI_2+a\sigma_1+b\sigma_2+c\sigma_3)$ with $K, a, b, c\in \mathcal{R}$ and $a^2+b^2+c^2\not=0$.
It is obvious that $(W^{[K,a,b,c]})_{11}=\frac{K+c}{2}$ and
$(W^{[K,a,b,c]})_{22}=\frac{K-c}{2}$. Therefore, $\Delta=[\frac{K-|c|}{2},\frac{K+|c|}{2}]$.
Moreover, $Tr[W^{[K,a,b,c]}\rho]=\frac{1}{2}(K+ax+by+cz)\in \mathcal{R}\backslash\Delta$
is equivalent to that $|ax+by+cz|>|c|$.

Denote $\Pi_1:$ $ax+by+cz=|c|$ and $\Pi_2:$  $ax+by+cz=-|c|$. $\Pi_1$ and $\Pi_2$ define two planes in a three dimensional space. Let $d_{\Pi_i}$ be the distance from the coordinate origin $O(0,0,0)$ to the plane $\Pi_i$, $i=1,2$. It is clear that $d_{\Pi_1}=d_{\Pi_2}=\frac{|c|}{\sqrt{a^2+b^2+c^2}}$. By the Definition \ref{D1} $W^{[K,a,b,c]}$ is an effective coherent witness of some states $\rho$ if and only if
there are two intersections between the plane $\Pi_i$ $(i\in\{1,2\})$ and the spherical surface $x^2+y^2+z^2=1$. In addition, a sufficient and necessary condition to have such two intersections is $d_{\Pi_i}<1$, i.e., $|c|<\sqrt{a^2+b^2+c^2}$. We conclude by remarking that  $W^{[K,a,b,c]}$ is an effective coherent witness of some states $\rho$ if and only if $a^2+b^2\not=0.$ Taking $a=b=1$ and $c=1$ as an example, $W^{[K,1,1,1]}$ is an effective coherent witness for the states $\rho\in\{\frac{1}{2}(I_2+x\sigma_1+y\sigma_2+z\sigma_3)||x+y+z|>1\}$, see Fig. 1.
\begin{figure}[htpb]
\renewcommand{\figurename}{Fig.}
\centering
\includegraphics[width=7.5cm]{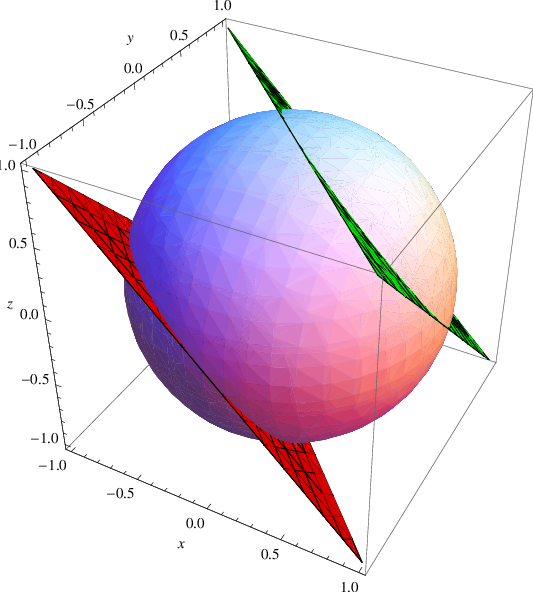}
\caption{{\small The ball is $x^2+y^2+z^2\leq1$, the red plane is $x+y+z=-1$ and the green plane is $x+y+z=1$.}}
\end{figure}

According to $d_{\Pi_i}$ $(i=1,2)$, we denote
\begin{equation*}
C(a_1,a_2,b_1,b_2)
=\left\{W^{[K,a_1,b_1,0]},W^{[K,a_2,b_2,0]}|a_1:b_1\not=a_2:b_2~and~ a^2_i+b^2_i\not=0~(i=1,2)\right\}.
\end{equation*}
It is obvious that $(W^{[K,a_i,b_i,0]})_{11}=(W^{[K,a_i,b_i,0]})_{22}=\frac{K}{2}$, i.e., $C(a_1,a_2,b_1,b_2)\in C^{2}_{[\frac{K}{2}, \frac{K}{2}]}$. It is seen that $W^{[K,a_i,b_i,0]}$ is an effective coherent witness for the states $\rho$ with $a_ix+b_iy\not=0$ since $Tr(W^{[K,a_i,b_i,0]}\rho)=\frac{K+a_ix+b_iy}{2}$, namely, $W^{[K,a_i,b_i,0]}$ cannot detect the coherence of the states on the plane $a_ix+b_iy=0$. However, the coherence of states on the plane $a_ix+b_iy=0$, except the states on the z-axis, can be detected by $W^{[K,a_j,b_j,0]}$ since the point $(x,y,z)\in\{(x,y,z)|a_ix+b_iy=0,~x^2+y^2\not=0\}$ satisfies that $a_jx+b_jy\not=0$ for $i\not=j\in\{1,2\}$. Hence, we have $[C(a_1,a_2,b_1,b_2)](\rho)=\{\rho|x^2+y^2\not=0\}$.
On the other hand, the states $\rho$ are incoherent when $x^2+y^2=0$. Consequently, we obtain that $[C(a_1,a_2,b_1,b_2)](\rho)=\mathcal{D} \setminus \mathcal{I}$. Taking $a_1=a_2=b_1=1$ and $b_2=-1$ in $C(a_1,a_2,b_1,b_2)$, we have the Fig. 2.
\begin{figure}[htpb]
\renewcommand{\figurename}{Fig.}
\centering
\includegraphics[width=7.5cm]{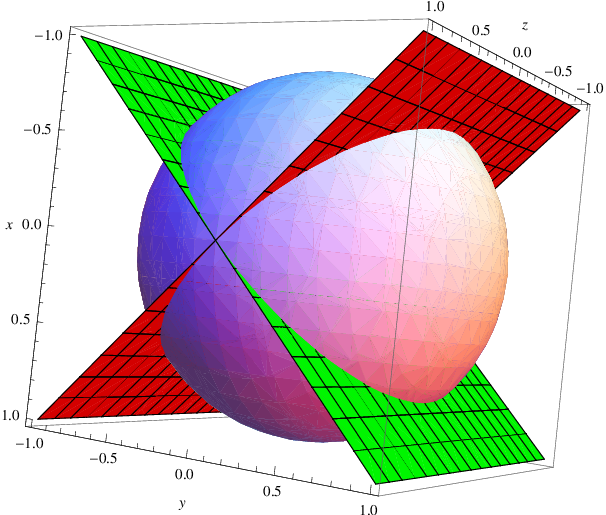}
\caption{{\small The ball is $x^2+y^2+z^2\leq1$, the red plane is $x+y=0$, and  the green plane is $x-y=0$.}}
\end{figure}

From Example 1, we have that all qubit coherent states can be detected by a finite set of coherence witnesses. Next let us consider $d$-dimensional $(d\geq2)$ case. The quantum states are given by density operators \cite{pp},
\begin{equation}\label{p}
\rho=\frac{1}{d}(I_d+\sum_{i=1}^{d^2-1}r_i\lambda_i),
\end{equation}
where $\lambda_i$, $i=1,2,...,d^2-1$, are the generators of $SU(d)$ given by $\{\omega_l,u_{jk},v_{jk}\}$ with
$\omega_l=\sqrt{\frac{2}{(l+1)(l+2)}}\left(\sum_{i=0}^{l}|i\rangle\langle i|-(l+1)|l+1\rangle\langle l+1|\right)$, $u_{jk}=|j\rangle\langle k|+|k\rangle\langle j|$ and $v_{jk}=-i(|j\rangle\langle k|-|k\rangle\langle j|)$, $0\leq l\leq d-2$, $0\leq j< k\leq d-1$ and $r_i=\frac{d}{2}Tr(\rho\lambda_i)$. Denote the Bloch vector
$\vec{r}=(r_1,...,r_{d^2-1})^t \in \mathcal{R}^{d^2-1}$, where $t$ stands for transpose.
One has \cite{r} $||\vec{r}||_2=\sqrt{\sum_{i=1}^{d^2-1}r^2_i}\leq \sqrt{\frac{d(d-1)}{2}}$, where $||\cdot||_2$ is the Euclidean norm on $\mathcal{R}^{d^2-1}$.

Without loss of generality, let
\begin{equation}\label{y}
\lambda_i=\left\{
\begin{aligned}
&~~~ \omega_{i-1},~~~ ~~~~~~ 1\leq i\leq d-1;\\
&~~~u_{jk},~~~~~~~~~d\leq i\leq \frac{1}{2}(d-1)(d+2)~and~0\leq j<k\leq d-1;\\
&~~~v_{jk},~~~~~~~~~ \frac{1}{2}d(d+1)\leq i\leq d^2-1~and~0\leq j<k\leq d-1.
\end{aligned}
\right.
\end{equation}
Denote
\begin{equation}\label{Wa}
W_{\vec{\eta}}=
\frac{1}{d}\big(KI_d+\sum_{i=1}^{d^2-1}s_i\lambda_i\big)
\end{equation}
and $\vec{\eta}=\big(s_1,s_2,...,s_{d^2-1}\big)^t\in\mathcal{R}^{d^2-1}$.
Let $\vec{T}_{ij}=\big(\langle i|\lambda_1|j\rangle, \langle i|\lambda_2|j\rangle,...,\langle i|\lambda_{d^2-1}|j\rangle\big)^t$. According to (\ref{y}), we have
\begin{equation*}
\vec{T}_{ij}=
\left\{
\begin{aligned}
&~ \left(\langle i|\lambda_1|i\rangle),...,\langle i|\lambda_{d-1}|i\rangle,0,...,0\right)^t,~~~~i=j;\\
&~\left(0,..,0,\langle i|\lambda_d|j\rangle),...,\langle i|\lambda_{d^2-1}|j\rangle\right)^t,~~~i\not=j.
\end{aligned}
\right.
\end{equation*}

\begin{theorem}\label{TH2}
For each coherent state $\rho$ of $d$ $(d\geq2)$ level quantum systems,
there is a witness $W_{\vec{\eta}}\in C^{d}_{\vec{\eta}} = \{W_{\vec{\eta}}|s_i=0~( 1\leq i\leq d-1)~for ~all ~\vec{\eta}\}$ which detects the coherence of $\rho$, i.e., $C^{d}_{\vec{\eta}}(\rho)=\mathcal{D} \setminus \mathcal{I}$.
\end{theorem}

{\sf [Proof]}~Since $W_{\vec{\eta}}=[(W_{\vec{\eta}})_{ij}]_{d\times d}$ with $\vec{\eta}=(0,0,...,0,s_{d},...,s_{d^2-1})$ has the form (\ref{Wa}), it is obvious that
\begin{eqnarray*}
W_{pp}&=&\langle p|W_{\vec{\eta}}|p\rangle\\
&=&\frac{1}{d}\left(K+\sum^{d^2-1}_{i=1}s_i\langle p|\lambda_i|p\rangle\right)\\
&=&\frac{1}{d}
\left(K+(\vec{\eta}_d)^t\vec{T}_{pp}\right)=\frac{K}{d}
\end{eqnarray*}
for all $p=1,2,...,d.$
According to Theorem \ref{TH1}, we need to prove that there exists a
$W_{\vec{\eta}}\in C^{d}_{\vec{\eta}}$ such that $Tr(W_{\vec{\eta}}\,\rho)\not=\frac{K}{d}$ for any given coherent state $\rho$.

Given each coherent state $\rho\in\mathcal{D} \setminus \mathcal{I}$, without loss of
generality, we assume $\rho_{ij}\not=0$ for some $i\not= j$, $\{i,j\}\in\{1,2,...,d\}$.
Note that $\rho_{ij}=\langle i|\rho|j\rangle
=\frac{1}{d}\sum_{m=1}^{d^2-1}r_m\langle i|\lambda_m|j\rangle
=\frac{1}{d}(\vec{r})^t\vec{T}_{ij}\not=0$ implies that
there must be a $m_0\in\{d,..,d^2-1\}$ such that $r_{m_0}\not=0$, since $\vec{T}_{ij}=\left(0,..,0,\langle i|\lambda_d|j\rangle),...,\langle i|\lambda_{d^2-1}|j\rangle\right)^t$.

Let us take the vector $\vec{\eta}^{m_0}=(0,...,0,1,0...,0)^t$ whose elements are zero except for $s_{m_0}=1$, i.e., $W_{\vec{\eta}^{m_0}}\in C^{d}_{\vec{\eta}}.$
We easily verify that $Tr(W_{\vec{\eta}^{m_0}}\rho)
=\frac{K}{d}+\frac{2}{d^2}
(\vec{r})^t\vec{\eta}^{m_0}=\frac{K}{d}+\frac{2r_{m_0}}{d^2}\not
=\frac{K}{d}$. Therefore, $W_{\vec{\eta}^{m_0}}$ is the coherence witness of $\rho$ and
$C^{d}_{\vec{\eta}}(\rho)=\mathcal{D} \setminus \mathcal{I}$. $\Box$

For the qubit systems, we have $\lambda_1=\sigma_3,$ $\lambda_2=\sigma_1$ and
$\lambda_3=\sigma_2$. According to Theorem \ref{TH2}, $\vec{\eta}=(0,a,b)^t$ and  $C^{2}_{\vec{\eta}}=\{W_{\vec{\eta}}=\frac{KI_2+a\sigma_1+b\sigma_2}{2}\}$. In fact, from Example 1 one may chose two coherence witnesses $W^{[K,a_1,b_1,0]}$ and $W^{[K,a_2,b_2,0]}$ to detect all the coherent states, where $a_1:b_1\not=a_2:b_2$ and $a^2_i+b^2_i\not=0$. Namely,
one can simplify $C^{2}_{\vec{\eta}}$ to be a finitely set $C(a_1,a_2,b_1,b_2)$.

A natural question is if the set $C^{d}_{\vec{\eta}}$ $(d\geq3)$ can be also simplified.
For each coherent state $\rho$ of $d$-level quantum systems, there must exist
$i\in\{d,d+1...,d^2-1\}$ such that $r_{i}\not=0$. Denote $P$ the set of all coherent states $\rho$ and $P_{i}=\{\rho|r_i\not=0\}$, $i=d,...,d^2-1$. It is obvious that $P_{d}\bigcup  P_{d+1} \bigcup...\bigcup P_{d^2-1}=P$. Let $\vec{\eta}^i$ be a vector whose elements are zero except for $s_{i}\not=0$, $i\in\{d,d+1,...,d^2-1\}$, i.e., $\vec{\eta}^i=(0,0,...,0,s_i,0,..,0)^t$. We have $Tr(W_{\vec{\eta}^i}\rho)
=\frac{K}{d}+\frac{2}{d^2}
(\vec{r})^t\vec{\eta}^i=\frac{K}{d}+\frac{2r_is_i}{d^2}\not
=\frac{K}{d}$ for all $\rho\in P_i,$ which shows that the witness $W_{\vec{\eta}^i}$
detects the coherence of all the states $\rho\in P_i$. In this way we have $d(d-1)$ coherence witnesses to detect all the coherent states $\rho$. Denote $\widetilde{C^{d}}[s_d,s_{d+1},...,s_{d^2-1}]=\{W_{\vec{\eta}^i}|
\vec{\eta}^i=(0,0,...,0,s_i,0,..,0)^t~with~s_i\not=0~(i=d,...,d^2-1)\}$.
We have the following corollary.

\begin{corollary}\label{c1}
All the coherent states $\rho$ of $d$ $(d\geq2)$ level quantum systems can be detected by a finite set of $d(d-1)$ witnesses $\widetilde{C^{d}}[s_d,s_{d+1},...,s_{d^2-1}]\subset C^{d}_{[\frac{K}{d},\frac{K}{d}]}$, i.e., $\widetilde{C^{d}}[s_d,s_{d+1},...,s_{d^2-1}](\rho)=\mathcal{D} \setminus \mathcal{I}$.
\end{corollary}

According to Corollary \ref{c1}, we have $\widetilde{C^{2}}[s_2,s_3]
=\{\frac{KI_2+s_2\sigma_1}{2},\frac{KI_2+s_3\sigma_2}{2}|s_2\not=0~ and~s_3\not=0\}$ for qubit systems. It is obvious that $\widetilde{C^{2}}[s_2,s_3]=C(a_1,a_2,b_1,b_2)$ with
$a_1=s_2$, $b_1=0$, $a_2=0$ and $b_2=s_3$.

It is seen that the ability of the coherence witness $W_{\vec{\eta}}$ in detecting coherent states is independent of the parameter $K$. Based on this fact, we have presented $C^{d}_{\vec{\eta}}(\rho)=\mathcal{D} \setminus \mathcal{I}$ no matter what the value $K$ takes in $W_{\vec{\eta}}$. Moreover, since $|(\vec{r})^t\vec{\eta}| \leq||\vec{r}||_2||\vec{\eta}||_2
\leq\sqrt{\frac{d(d-1)}{2}}||\vec{\eta}||_2$, for any given $\rho\in W_{\vec{\eta}}(\rho)$ we have $Tr(W_{\vec{\eta}}\,\rho)=
\frac{Kd+2(\vec{r})^t\vec{\eta}}{d^2}
\in\left(\frac{Kd-\sqrt{2d(d-1)}||\vec{\eta}||_2}{d^2},\frac{K}{d}\right)
\bigcup\left(\frac{K}{d},
\frac{Kd+\sqrt{2d(d-1)}||\vec{\eta}||_2}{d^2}\right)$.
In particular, if $W_{\vec{\eta}}\in\widetilde{C^{d}}[s_d,s_{d+1},...,s_{d^2-1}]$  and $\rho\in W_{\vec{\eta}}(\rho)$, we have $Tr(W_{\vec{\eta}}\,\rho)
\in\left(\frac{Kd-\sqrt{2d(d-1)}|S|}{d^2},\frac{K}{d}\right)
\bigcup \left(\frac{K}{d},
\frac{Kd+\sqrt{2d(d-1)}|S|}{d^2}\right)$, where $|S|=\max\{|s_d|,|s_{d+1}|,...,|s_{d^2-1}|\}$.

\section{Conclusion and discussion}
By introducing the useful concept of coherence witness $W$, we have derived explicitly the sufficient and necessary conditions for the detection of the coherence of any state $\rho$. In particular, for qubit systems, we have provided detailed construction of coherence witnesses $W^{[K,a,b,c]}$, and calculated the finite number of coherence witnesses that detect all the coherent states of qubit systems.

For general $d-$dimensional quantum systems, based on the Bloch vector expansion of density matrices, we have shown that each coherent state can be detected by one coherence witness of the set $C^{d}_{\vec{\eta}}$. Moreover, we have classified the states into $d(d-1)$ categories: $P_i$, $i=d,d+1,...,d^2-1$, according to Bloch vector $\vec{r}$. We have constructed a coherence witness $W_{\vec{\eta}^i}$ to detect the coherence of $\rho\in P_i$ for any $i=d,d+1,...,d^2-1$. These $d(d-1)$ coherence witnesses detect all the coherent states.
It is noteworthy that we only consider the  finite dimensional quantum systems in this paper.
Our approach may be also shed light on the investigations on coherence in multipartite finite dimensional systems.

\bigskip
\noindent{\bf Acknowledgments}\, \,
This work is supported by the National Natural Science Foundation of China under
grant Nos. 12075159, 12171044 and 12301582; the specific research fund of the Innovation Platform for Academicians of Hainan Province under Grant No. YSPTZX202215; Guangdong Basic and Applied Basic Research Foundation under Grant No. 2024A1515030023; and the Start-up Funding of Dongguan University of Technology No. 221110084.

\end{document}